\documentclass[twocolumn,showpacs,preprintnumbers]{revtex4}
\usepackage{amssymb}
\usepackage{amsmath}
\usepackage{graphicx}
\usepackage{epsfig}

\begin{document}
\title{\bf Evidence for two-gap superconductivity in Ba$_{0.55}$K$_{0.45}$Fe$_2$As$_2$ by directional point-contact Andreev-reflection spectroscopy.}

\author{P. Szab\'o,$^{1}$ Z. Pribulov\'a,$^{1}$ G. Prist\'a\v s,$^{1}$ 
 S. L. Bud'ko,$^{2}$ P. C. Canfield,$^{2}$ and P. Samuely$^{1}$ }

\affiliation{$^1$Centre of  Low Temperature Physics, Institute of Experimental Physics, Slovak
Academy of Sciences\& P.J. \v Saf\'arik University, Watsonova
47, SK-04001 Ko\v sice, Slovakia\\
$^2$Ames Laboratory, Iowa State University, Ames, IA 50011, USA.
}

\begin{abstract}

Directional point-contact Andreev-reflection spectroscopy measurements on the Ba$_{0.55}$K$_{0.45}$Fe$_2$As$_2$ single crystals are presented. The spectra show significant differences when measured in the $ab$ plane in comparison with those measured in the $c$ direction of the crystal. In the latter case  only a reduced point-contact conductance around zero bias has been revealed  persisting well above $T_c$ and probably related to the structural and magnetic transitions in the system. Within the $ab$ plane two superconducting energy gaps are detected below $T_c$. Here a reduced conductance above $T_c$ could also be found. The fits of the $ab$-plane data to the superconducting $s$-wave two-gap model indicate that the smaller gap has a size below the BCS value while the large gap reveals much higher coupling strength.

\end{abstract}

\maketitle


REFeAsO(F) systems with various rare earth (RE) elements  bring a new class of layered high-$T_c$ materials without copper \cite{kamihara}. NdFeAsO$_{0.9}$F$_{0.1}$ with $T_c$ above 51 K  \cite{ren} together with the Sm and Pr compounds reveal the highest transition temperature at ambient pressure among all oxypnictides. Besides REFeAsO(F), another three groups within pnictides have been identified. Namely, \textit{A}Fe$_2$As$_2$ with \textit{A} = Ba, Sr and Ca made superconducting by chemical substitution \cite{rotter} or pressure \cite{torikachvili}, Li$_x$FeAs \cite{wang} and $\alpha$-FeSe \cite{hsu} are intensively studied.

One of the important questions to be addressed in these multiband systems concerns the symmetry of the superconducting order parameter(s). There are already many theoretical predictions  on this topic but also a body of  experimental studies is emerging. Band structure calculations have shown disconnected sheets of Fermi surfaces with possibly different superconducting energy gaps. A minimal model has to include two bands: the hole band around the $\Gamma$ point and the electron one around the $M$ point \cite{scalapino}. In contrast to the multiband but conventional $s$-wave scenario in MgB$_2$ \cite{szabo}, here, the extending $s$-wave pairing with a sign reversal of the order parameter between different Fermi surface sheets was proposed by Mazin {\it et al.} \cite{mazin}. Experimental results are providing controversial conclusions.
NMR studies on PrFeAsO(F) \cite{matano} found two superconducting energy gaps with nodes. Also very recent muon spin relaxation measurements on single crystalline Ba$_{0.5}$K$_{0.5}$Fe$_2$As$_2$ \cite{goko} point to the gap with nodes. On the other hand the ARPES measurements \cite{ding} have found two nodeless and nearly isotropic gaps in Ba$_{0.6}$K$_{0.4}$Fe$_2$As$_2$   and the penetration depth studies have provided the nodeless superconducting energy gap in NdFeAsO$_{0.9}$F$_{0.1}$ but with remarkably small coupling 2$\Delta/k_BT_c \approx $ 2 \cite{martin}. 

By the point-contact Andreev reflection (PCAR) spectroscopy: Wang {\it et al.} \cite{shan} found a nodal superconductivity with multiple gaps in SmFeAsO(F). Chen {\it et al.} \cite{tesanovic} presented a surprisingly conventional superconducting energy gap with a medium coupling equal to 2$\Delta/k_BT_c \approx 3.7$. The recent data of Yates {\it et al.} \cite{cohen} obtained on the 45 K NdFeAsO$_{0.85}$ show also an indication of the superconducting energy gap with 2$\Delta/k_BT_c = 3.6$. 
In our previous paper \cite{samuely} two nodeless gaps have been found by the PCAR measurements on the polycrystalline samples of NdFeAsO(F).

In the following we present directional PCAR  studies on iron pnictides, namely on the Ba$_{0.55}$K$_{0.45}$Fe$_2$As$_2$ single crystals. 
The spectra have shown significant differences when measured in the $ab$ plane and in the $c$ direction. In the latter case no suitable conditions to observe superconducting energy gaps with coherent gap-like peaks and  enhanced conductance in the PCAR spectrum have been found. A predominant feature has been revealed in the form of a reduced point-contact conductance  persisting well above $T_c$ which can be attributed to a reduced density of states (DOS) due to a structural phase transition \cite{ni} and magnetism found in a partial volume fraction in these  samples  below 70 K \cite{goko}. On the other hand within the $ab$ plane two  superconducting energy gaps are found. The pair of peaks due to a small gap has been scattered  between $ \pm(3-5)$ mV while the humps indicating the presence of the second gap are placed at  $\pm(10-12)$ mV. In the superconducting gap-like spectra the reduced conductance above $T_c$ could also be found. The $ab$-plane  spectra have been successfully fitted to the $s$-wave two-gap model providing the conventional temperature dependence of the two gaps with the coupling strengths $2\Delta_1 /kT_c = 2.5-4$ and $2\Delta_2/kT_c = 9-10$.

The Ba$_{0.55}$K$_{0.45}$Fe$_2$As$_2$ single crystals were grown out of a Sn flux as described in details in Ref.\cite{ni}, where the following characteristics have been found:
typical dimensions of the crystals are 2x2x0.1 mm$^3$. The crystallographic $c$ axis is perpendicular to the plane of the plate-like crystals. The resistive measurements have shown the onset of the superconducting transitions below 30 K and the  zero resistance  at 27 K. Rather broad transitions in some of the crystals with multiple steps  are attributed to a possible different amount of potassium in different layers or different crystals. The specific heat as well as the resistivity measurements on these crystals has shown features at about 85 K. Although reduced they are found at the same temperature as on the undoped BaFe$_2$As$_2$ samples where the tetragonal-to-orthorhombic structural phase transition takes place. This transition associated also with a reduction in the density of states at the Fermi level (indicated by the Hall effect) has been revealed here at lower temperature as compared with 140 K found in Ref.\cite{rotter}. The decreased structural transition temperature is due to the
amounts of Sn up to 1\% incorporated into the bulk. A presence of tin does not significantly effect the high-temperature
superconducting phase transition in
B$_{0.55}$K$_{0.45}$Fe$_2$As$_2$.
In the present work a local transition temperature has been measured by the point-contact technique showing superconducting $T_c$'s between 23 and 27 K.
 
The crystals were cleaved to reveal fresh surface before point-contact measurements. 
For the measurements in the $c$ direction  the fresh shiny surface has been obtained by detaching the degraded surface layers  by a scotch tape. The  microconstrictions  were prepared {\it in situ}  by pressing  a metallic  tip  (platinum wire formed either  mechanically or  by electrochemical etching) on a fresh surface
of  the  superconductor.  For the measurements with the point-contact current in the $ab$ plane usually a reversed tip-sample configuration has been used. The freshly cleaved edge of the single crystal jetting out in $ab$ direction has been pressed on a piece of chemically etched copper.  A special point-contact approaching system allowed for
lateral as well as vertical movements of the point-contact tip by a differential
screw  mechanism. Details of the technique can be found elsewhere \cite{szabo}.

PCAR  spectra measured    on   the    ballistic
microconstriction   between   a   normal   metal  and
a superconductor consists of  pure AR and tunneling contributions, respectively \cite{btk}.
First contribution makes the conductance inside $|V| < \Delta/e $ twice as large as in the normal state or as what is at large bias  where the coupling via  the gap $\Delta$ is inefficient. Tunneling contribution 
reduces the  conductance at  the zero  bias and two
symmetrically  located peaks  rise  at  the gap  energy. PCAR
conductance can be  compared with  the Blonder-Tinkham-Klapwijk (BTK) model
using  as input parameters  the energy gap  $\Delta$,
the parameter $z$ (measure for the strength of the interface
barrier)   and  a   parameter  $\Gamma$   for  the  spectral
broadening \cite{plecenik}. For   a  multiband/multigap superconductor the
point-contact conductance $G = dI/dV$ can be expressed as a weighted
sum of  partial BTK  conductances. As shown in our previous work \cite{szabo} on the two-gap superconductor of MgB$_2$ the total PCAR conductance  consists of two parallel contributions, the first originated  from a three-dimensional (3D)  $\pi$ band with a small
gap $\Delta_{1}$ and a weight $\alpha$ and the second from a  quasi-two-dimensional  $\sigma$ band  with  a  large  gap $\Delta_{2}$, respectively.

Spectral characteristics as the superconducting energy gaps or electron-phonon interaction features can be read from the point-contact data only if the junction is in a ballistic or diffusive regime, where heating effects are avoided. There, both the elastic ($l_e$) as well as inelastic ($l_i$) mean-free paths  should be bigger than the diameter $d$ of the junction or a diffusion length $\sqrt{l_el_i} > d$. These requirements should be satisfied in the normal as well as in the superconducting part of the junction \cite{btk}. The point contacts made on the Ba$_{0.55}$K$_{0.45}$Fe$_2$As$_2$ single crystals have resistances between tens and hundreds of ohms which corresponds to the contact diameter of tens of nanometers as calculated from the Wexler's formula \cite{naidyuk} for the point contact on a highly resistive material as  Ba$_{0.55}$K$_{0.45}$Fe$_2$As$_2$ \cite{ni}. Indeed, the electron mean-free path, although not known in Ba$_{0.55}$K$_{0.45}$Fe$_2$As$_2$ could also be  on the order of tens of nanometers.  Then, the precautions should be made to avoid the junctions with heating effects. In the following only the junctions without the conductance dips and irreversibilities in voltage dependences are presented. Predominantly the junctions with a finite barrier-strength parameter $z$ allowing for tunneling component in the spectrum were examined.

\begin{figure}[t]
\includegraphics[width=7.6 cm]{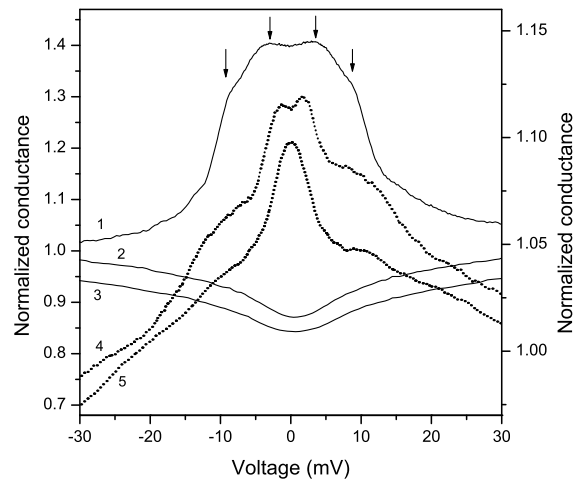}
\caption{Typical PCAR spectra measured with the PC current mostly within the $ab$ plane of the Ba$_{0.55}$K$_{0.45}$Fe$_2$As$_2$ crystal showing two superconducting gaps indicated by arrows (curves 1, 4, and 5) and in the $c$-direction showing reduced PC conductance near zero bias (curves 2, and 3). For the curves 4 and 5 the right ordinate applies.}
\label{fig:fig1}
\end{figure}

\begin{figure}[t]
\includegraphics[width=7.6 cm]{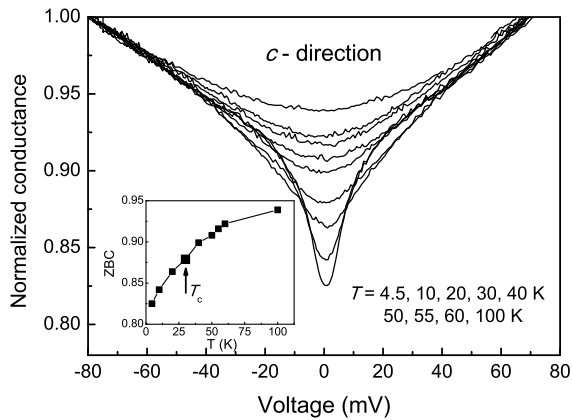}
\caption{Spectrum measured in the $c$ direction showing a reduced conductance even above $T_c$.}
\label{fig:fig2}
\end{figure}

Figure 1 shows typical PCAR spectra obtained on Pt-Ba$_{0.55}$K$_{0.45}$Fe$_2$As$_2$ point-contacts with the current preferably within the $ab$ plane (curves 1, 4, and 5) and in the $c$ direction (curves 2, and 3) of the crystal. All  spectra shown here are raw data, just normalized to their own values at $V$ = 50 mV eventually vertically shifted for the clarity. The spectra 1, 4 and 5 present (double) enhanced conductances, the typical features of the Andreev reflection of quasiparticles coupled via the superconducting energy gap(s). The first enhancement starts below 20 mV with the gap-like humps at about $\pm 10$ mV while the second one is located below $~ 5-7$ mV. On the spectra 1 and 4 also two symmetrical maxima at $\pm$4 and $\pm 2$ mV, respectively, are displayed. Majority of  tens of the spectra measured in the $ab$ direction revealed a heavily broadened enhanced conductance near the zero bias as indicated by the spectrum 5. 

The measurements with the point-contact current along the $c$ axis yield a completely different picture. For this direction we have collected  tens of spectra without any enhanced PCAR conductance and coherence peaks due to superconducting energy gap(s). On the contrary the spectra display  a reduced conductance around the zero-bias conductance (ZBC). In Fig. 2 one can find the temperature dependence of such a spectrum. The reduced conductance is gradually smeared and filled up when increasing the temperature. Moreover,  as could be seen from the measurements at higher temperatures and from the ZBC vs. temperature dependence (inset of Fig. 2) the effect of the reduced conductance persists well above the superconducting transition up to about the N\'eel temperature $T_N \approx $ 70 K found for a similar sample \cite{goko}. Thus, within the $c$ direction suitable spectral conditions for observing the superconducting energy gap have not been found, possibly due to a surface reconstruction or contamination.  Then, the observed reduced conductance can be just a differential-conductance-versus-voltage replica of the increasing resistivity with decreasing temperature below 85 K found in the undoped sample \cite{ni}. Another possibility is that it is a spectral characteristic where the superconducting gap features are heavily smeared.

\begin{figure}[t]
\includegraphics[width=7.6 cm]{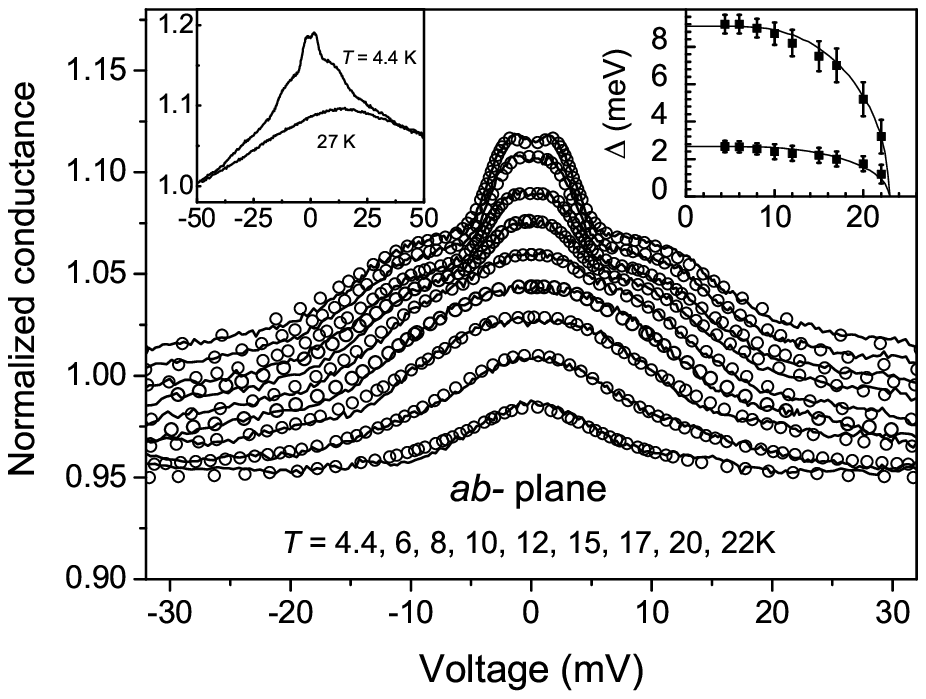}
\caption{Normalized PCAR spectra of the Ba$_{0.55}$K$_{0.45}$Fe$_2$As$_2$ single crystal in the $ab$ plane showing two superconducting gaps. Lower curves are vertically shifted for clarity. Circles represent the best fits to the two gap BTK formula. Left inset -  raw data taken at 4.4 and 27 K, right inset - temperature dependence of two gaps, solid lines show BCS-type behavior of energy gaps.}
\label{fig:fig3}
\end{figure}

Figure 3 shows the representative spectra measured within the $ab$ planes at different temperatures. 
The left inset presents the raw data taken at 4.4 and 27 K. As can be seen the spectral background reveals asymmetry, being higher at positive bias voltage, i.e., at negative sample bias.  This asymmetry seems to have different sign for the 122 pnictides (see also \cite{boyer} for Sr$_{1-x}$K$_x$Fe$_2$As$_2$) than  and for the 1111 superconductors as indicated by Chen {\it et al.} \cite{tesanovic} or in our previous measurements on the polycrystalline  NdFeAsO(F) \cite{samuely}.
For a better comparison with the BTK model the spectra in the main figure are normalized to the normal-state curve measured at 27 K. The remarkable feature at the lowest temperature is a double enhanced point-contact conductance. The form of the spectrum obviously reminds the  two-gap spectrum of MgB$_2$ for a highly transparent junction with conductance enhancements due to Andreev reflection of quasiparticles on the small and large superconducting energy gaps. With increasing temperature the two-gap structure is smoothly smeared out. Indeed, the spectra from Fig. 3 could be easily fitted to the two-gap BTK formula. The best fit for each temperature is shown by open circles. The right inset of Fig. 3 shows the temperature dependence of the superconducting energy gaps resulting from the fitting procedure. For comparison we also show by solid lines the BCS-type temperature dependence of $\Delta$'s. The values of the energy gaps at lowest temperature are for the small $\Delta_1 \approx $ 2.7 meV and the large one $\Delta_2 \approx $ 9.2 meV, which corresponds to the coupling strengths $2\Delta_1 /kT_c = 2.7$ and $2\Delta_2/kT_c = 9.3$ for $T_c=23$ K. The smearing parameters $\Gamma_1=1.8$ meV and $\Gamma_2=6$ meV, the barrier strengths $z_1=0.34$ and    $z_2=0.5$ as well as the weight factor  $\alpha=0.5$ obtained at 4.4 K were kept constant  at higher temperatures ($\Gamma 's$ changing less than 10 \%). 
From the  data obtained on many junctions we observe that the gaps are scattered as $2\Delta_1 /kT_c = 2.5-4$ and $2\Delta_2/kT_c = 9-10$.

Although the $s$-wave two-gap BTK formula has been successfully used to fit our PCAR data a possibility of unconventional pairing symmetry cannot be completely ruled out. Obviously,  rather strongly broadened spectra as presented here could be in principle fitted also  by anisotropic or nodal gaps, if an appropriate current injecting angle was selected \cite{tanaka}. On the other hand in no case the ZBC peak in the low temperature PCAR conductance as a fingerprint of the Andreev bound states due to $d$ or $p$-wave symmetry has been observed.

\begin{figure}[t]
\includegraphics[width=7.6 cm]{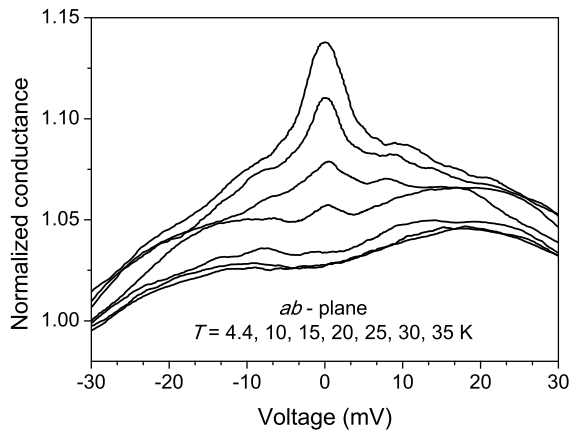}
\caption{PCAR spectra of the Ba$_{0.55}$K$_{0.45}$Fe$_2$As$_2$ single crystal showing the gap as well as the pseudogap features.}
\label{fig:fig4}
\end{figure}

Figure 4 presents another point-contact measured in the $ab$ plane documenting that besides the broadened Andreev-type enhancements of the conductance also a reduced conductance background around the zero bias is revealed on the same junction. The latter feature detectable near $T_c$ has persisted above $T_c$ similarly to the case of the $c$-axis 
spectra. This indicates that the reduced conductance is a spectral feature related to a reduced DOS in the normal state. 

Does the simultaneous occurrence of the superconducting energy gaps and reduced conductance on the same junction mean that these  two effects are due to quasiparticles from the same Fermi surface? 
Goko {\it et al.} \cite{goko} brought evidence  for  coexistence of superconductivity and strong static magnetic order in a partial volume fraction of (Ba,K)Fe$_2$As$_2$ and similar systems. Recently Park {\it et al.} \cite{park} found a mesoscopic phase separation on the single crystals of (Ba,K)Fe$_2$As$_2$  with  a similar $T_c \approx 32$ K as in our case.   This phase separation  leads to antiferromagnetically ordered and non-magnetic/superconducting regions, both with characteristic sizes of several tens of nanometers. Our point contacts have about the same size. Then, even if the reduced point-contact conductance is observed on the same junction which reveals also the superconducting energy gaps one has to admit that these two effects are caused by two spatially separated phases and not by the quasiparticles originating from the same Fermi surface. To resolve the problem of a possible coexistence of superconductivity and magnetism in a single electronic system the spectroscopic studies on the system without a mesoscopic phase separation are necessary.  

Very recently Gonnelli {\it et al.} \cite {gonnelli} also obtained pronounced two-gap spectra on the LaFeAsO$_{1-x}$F$_{x}$ polycrystals as well as  shown an existence of the reduced DOS in the normal state up to about 140 K, close to the N\'eel temperature of the undoped specimen. Their size of the large and small gaps are remarkably similar to those found in the present work on the 122 sample with a similar $T_c$. 

In conclusion, directional PCAR studies have been performed on the single crystalline   Ba$_{0.55}$K$_{0.45}$Fe$_2$As$_2$. Significant differences in the spectra are observed when measured within the $ab$ planes and in the $c$ direction. Within the planes two superconducting gaps  are detected with $2\Delta_1 /kT_c = 2.5-4$ and $2\Delta_2/kT_c = 9-10$. Within the $c$ direction only a reduced conductance could be found persisting well above $T_c$, possibly connected  with the reduced DOS in a mesoscopic antiferromagnetic region spatially separated from the superconducting phase.

This  work  was   supported  by  the Slovak R\&D Agency  under Contracts No. VVCE-0058-07, No. APVV-0346-07, and No. LPP-0101-06, the EC Framework Programme Grant No.MTKD-CT-2005-030002, and by the  U.S. Steel Ko\v sice, s.r.o.  Centre of Low Temperature Physics is operated as  the Centre of Excellence  of the Slovak Academy of Sciences.  
Work at the Ames Laboratory was supported by the Basic Energy Sciences, U.S. Department  of Energy, under  Contract No. DE-AC02-07CH11358. Valuable discussions with I.I. Mazin and N.L. Wang are appreciated.

\end{document}